\documentstyle[epsfig,12pt]{article}

\newskip\humongous \humongous=0pt plus 1000pt minus 1000pt

\newif\ifdtup

\begin{document}
\begin{titlepage}
\begin{center}
\today     \hfill    LBNL-41104 \\
~{} \hfill UCB-PTH-97/57  \\
\hfill SNS/PH/1997-8\\
\hfill LAL 97-91

\vskip .25in

{\large \bf Alternative Theories of CP violation}\footnote{This
work was supported in part by the Director, Office of
Energy Research, Office of High Energy and Nuclear Physics, Division of
High Energy Physics of the U.S. Department of Energy under Contract
DE-AC03-76SF00098 and in part by the National Science Foundation under
grant PHY-95-14797.}

%\footnote{This work was supported by the Director, Office of Energy
%Research, Office of High Energy and Nuclear Physics, Division of High
%Energy Physics of the U.S. Department of Energy under Contract
%DE-AC03-76SF00098.}
%
%alternate footnote for faculty:

\vskip 0.3in

Riccardo Barbieri$^1$, Lawrence Hall$^2$,
Achille Stocchi$^3$, and Neal Weiner$^2$

\vskip 0.1in
{{}$^1$ \em Scuola Normale Superiore, Pisa, Italy;\\
and INFN, Sezione di Pisa, Italy\\}

\vskip 0.1in

{{}$^2$ \em Department of Physics and
Lawrence Berkeley National Laboratory\\
University of California, Berkeley, California 94720\\}

\vskip 0.1in

{{}$^3$ \em Laboratoire de l'Acc\'el\'erateur Lin\'eaire  \\
   IN2P3-CNRS et Universit\'e de Paris-Sud, F-91405 Orsay }

\end{center}

\vskip .3in

\begin{abstract}

Recent improvements to the limit of $\Delta M_{B_s}$ imply that pure
superweak theories, while not excluded, no longer provide a good fit
to the data. A class of general superweak theories is introduced in
which all flavor changing interactions are governed by an approximate
flavor symmetry which gives a ``3 mechanism''. These theories are in
good agreement with data, and predict low values for $|V_{td}|,
|V_{ub}/V_{cb}|, B(K^+ \rightarrow \pi^+ \bar{\nu} \nu),
\epsilon'/\epsilon$ and CP asymmetries in B decays, and high values for
$\Delta M_{B_s}$ and $f_B \sqrt{B_B}$.
An important example of such a theory is provided by weak scale
supersymmetric theories with soft $CP$ violation. The $CP$ violation
originates in the squark mass matrix, and, with phases of order unity,
flavor symmetries can yield a correct prediction for the order of
magnitude of $\epsilon_K$.

\end{abstract}

\end{titlepage}

\section{CP Violation} \label{sec:cpv}

All observed CP violation can be described by the complex parameter
$\epsilon_K$, which describes an imaginary contribution to the $\Delta S = 2$
mixing of the neutral $K$ mesons. Such a mixing implies the existence of an
effective Hamiltonian
\begin{equation}
{\cal H}^{\Delta S = 2}_{eff} = {1 \over v^2} \sum_{ij} i C_{ij}
(\bar{s}\Gamma_i d) \; (\bar{s}\Gamma_j d)
\label{eq:ham}
\end{equation}
where $v = 247$ GeV, and $i,j$ run over possible gamma matrix structures.
The dimensionless coefficients $C_{ij}$ are real in a basis where the
standard model $\Delta S = 1$ effective Hamiltonian has a real coefficient.
In the case that the dominant term is $\Gamma_i = \Gamma_j = \gamma^\mu
(1-\gamma_5)/2$,
\begin{equation}
C_{LL} = 4(1 \pm 0.3) \cdot 10^{-10}  {|\epsilon_K| \over 2.3
\cdot 10^{-3}} {0.75 \over B_K}.
\label{eq:C}
\end{equation}

The two basic issues of CP violation are
\begin{itemize}
\item What is the underlying physics which leads to
${\cal H}^{\Delta S = 2}_{eff}$? Is it a very small effect originating at the
weak scale, as suggested by the form $C/v^2$, or is it a larger effect
generated by physics at higher energies?
\item How can the magnitude $C \approx 10^{-9}$ to $10^{-10}$ be
understood?
\end{itemize}

\section{The CKM Theory of CP Violation} \label{sec:ckm}

In the standard model all information about flavor and CP violation
originates from the Yukawa coupling matrices. After electroweak symmetry
breaking, this is manifested in the
Cabibbo-Kobayashi-Maskawa (CKM) matrix of the charged current
interactions of the $W$ boson \cite{ckm}.
A one loop box diagram with internal top quarks gives
the dominant contribution to ${\cal H}^{\Delta S = 2}_{eff}$ via
\begin{equation}
C_{LL,SM} = {g^2 \over 32 \pi^2} S_t {\mathcal I} m [(V_{td} V_{ts}^*)^2]
\label{eq:Csm}
\end{equation}
where $S_t \simeq 2.6$ is the result of the loop integration, and $g$ is the
SU(2) gauge coupling constant.
For a suitable choice of the CKM matrix elements,
$V_{ij}$, the standard model can provide a description of the observed CP
violation. The fundamental reason for the size of the CP violation observed
in nature remains a mystery, however, and must await a theory of flavor which
can explain the values of $|V_{td}|, |V_{ts}|$ and the CKM phase. If the CKM
matrix contained no small parameters one would expect $C_{LL,SM}$
to be of order $10^{-2}$ to $10^{-3}$ rather than the observed value of order
$10^{-9}$ to $10^{-10}$.

Of course, measurements of CP conserving observables have shown that
$|V_{ij}|$ are small for $i \neq j$, and,
given the measured values of $|V_{us}|$ and $|V_{cb}|$, it is
convenient to use the Wolfenstein parameterization\cite{Wp} of the CKM
matrix, in
which case (\ref{eq:Csm}) becomes
\begin{equation}
C_{LL,SM} \simeq 20 \cdot 10^{-10}  (1-\rho) \eta
\label{eq:Cwolf}
\end{equation}
If we {\it assume} that the CKM matrix
does not have any other small parameters, the standard model yields a value of
$\epsilon_K$ of the observed order of magnitude. While this is not a
prediction, it is an important success of the
standard model, and has made the CKM theory the leading candidate for CP
violation. To our knowledge, there is no similar success in any
published alternative to the CKM theory of CP violation, since in
these theories the order of magnitude of $C$ can only be fixed by
fitting to the measured value of $\epsilon_K$. In this letter we
present such an alternative theory.

Two further measurements of $|V_{ij}|$, with
$i \neq j$, would determine both $\rho$ and $\eta$ allowing a
prediction of $C_{LL,SM}$ and $\epsilon_K$.
A fit to the two observables
$|V_{ub}/V_{cb}|$ and $\Delta M_{B_d}$, but {\it not}
$\epsilon_K$, is shown in Figure 1.
For all numerical work, we use the data and parameters
listed in Table 1 --- for a discussion of these, and references, see
\cite{pprs}.
%we find $\rho = 0.055 \pm 0.25$** and
%$\eta = 0.36+0.13-0.22$.**
Unfortunately the large uncertainties make this
a very weak prediction: $\eta=0$ is allowed even at the 68\% confidence
level. Hence, from this one cannot claim strong evidence for CKM CP
violation.

Recent observations at LEP have improved the limit on $B_s -
\bar{B}_s$ mixing, so that $\Delta M_{B_s} > 10.2 \; \mbox{ps}^{-1}$ at
95\% confidence level \cite{jeru}. The result of a $\chi^2$ fit in the standard
model to $\rho$ and $\eta$ using the three observables
$|V_{ub}/V_{cb}|$, $\Delta M_{B_d}$ and $\Delta M_{B_s}$, but {\it not}
$\epsilon_K$, is shown in figure 2.
For $B_s$ mixing the amplitude method is used
\cite{mr,pprs}. Comparing Figures 1 and 2, it is clear that the
$\Delta M_{B_s}$ limit is now very significant.
At 68\% confidence level the standard model is able to predict the
value of $\epsilon_K$ to within a factor of 2; however, at 90\%
confidence level $\eta=0$ is allowed, so that at this level there is
no prediction, only an upper bound.
While this is an important success of the CKM theory, it is still worth
pursuing credible alternative theories of $CP$ violation.

\begin{figure}
\begin{center}
\epsfig{file=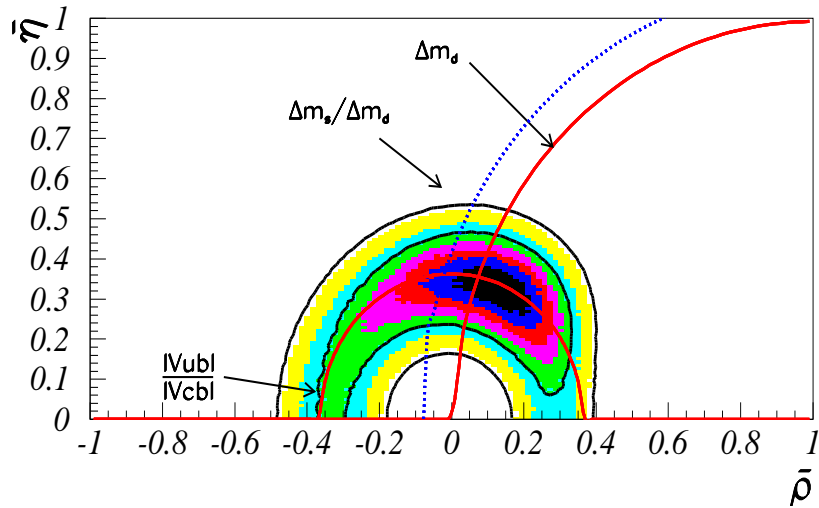, height=7cm, width=14cm,
bbllx=0cm, bblly=10cm, bburx=20cm, bbury=19cm}

\end{center}
\caption{The 68\% and 95\% C.L. contours fits of $|V_{ub}/V_{cb}|$ and
$\Delta M_{B_d}$ in the $\bar{\rho}/\bar{\eta}$ plane  in the standard model.
The curves correspond to constraints obtained from measurements of
$|V_{ub}/V_{cb}|, \Delta M_{B_d}$ and $\Delta M_{B_s}$ (The last constraint
is not included in the fit).
$\bar{\rho} = \rho (1 - \lambda^2/2), \bar{\eta} = \eta (1 - \lambda^2/2)$.}
\label{fig:1}
\end{figure}

\begin{figure}
\begin{center}
\epsfig{file=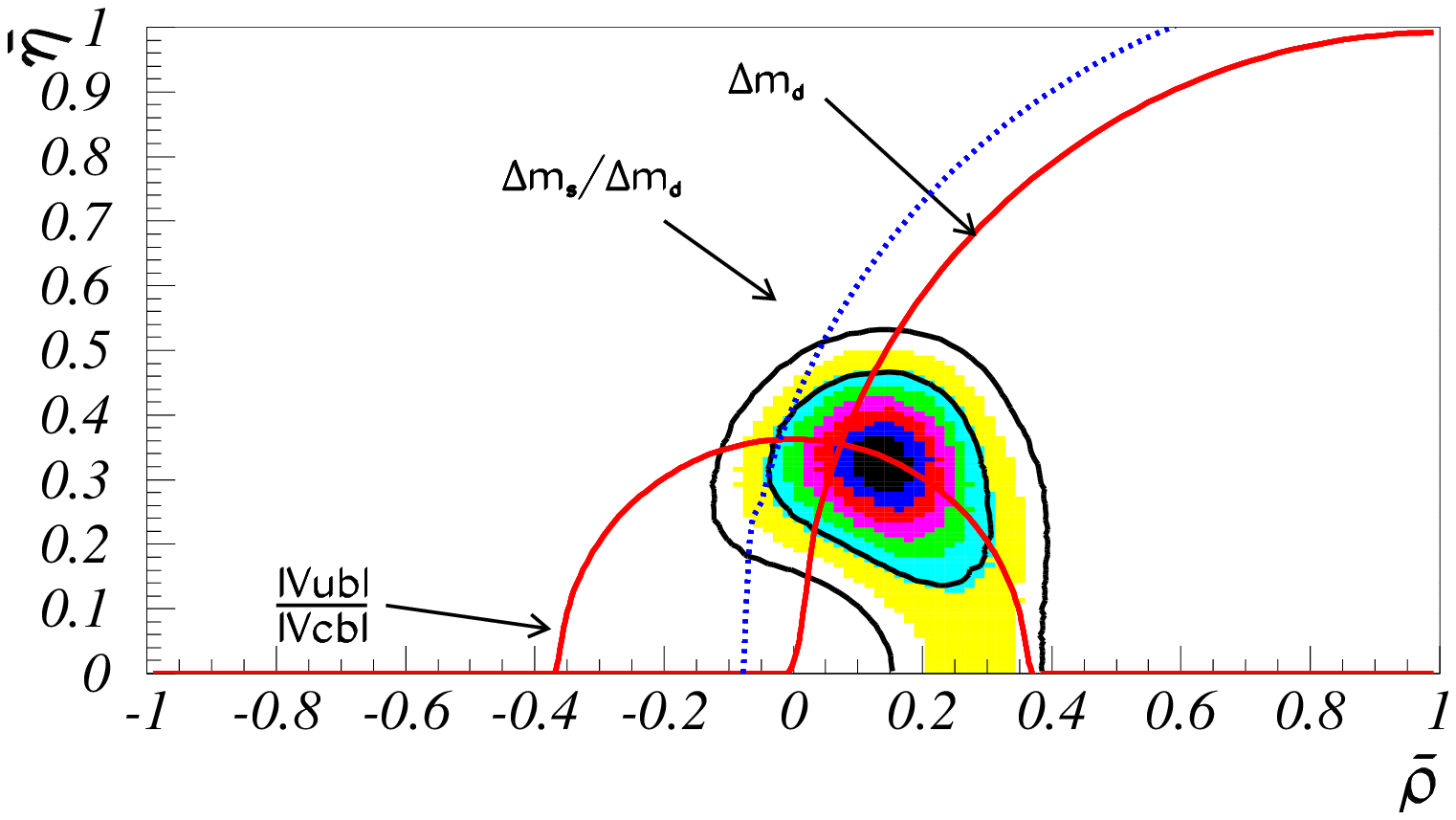, height=7cm, width=14cm,
bbllx=0cm, bblly=10cm, bburx=20cm, bbury=19cm}
\end{center}
\caption{The 68\% and 95\% C.L. contours fits of $|V_{ub}/V_{cb}|,
\Delta M_{B_d}$ and $\Delta M_{B_s}$ in the $\bar{\rho}/\bar{\eta}$ plane
in the standard model.
The curves correspond to constraints obtained from measurements of
$|V_{ub}/V_{cb}|, \Delta M_{B_d}$ and $\Delta M_{B_s}$.
$\bar{\rho} = \rho (1 - \lambda^2/2), \bar{\eta} = \eta (1 - \lambda^2/2)$.}
\label{fig:2}
\end{figure}

\begin{table}
\begin{center}
\caption{Values of observables and parameters}
\begin{tabular}{|c|c|}
\hline
$|V_{ub}/V_{cb}|$    & $0.080 \pm 0.020$ \\
\hline
$\Delta M_{B_d}$     & $0.472 \pm 0.018 \; \mbox{ps}^{-1}$ \\
\hline
$\Delta M_{B_s}$     & $>10.2 \mbox{ps}^{-1}$ at 95\% C.L.\\
\hline
$f_{B_d} \sqrt{B_{B_d}}$ &  $(200\pm 50)$ MeV  \\
\hline
$f_{B_s} \sqrt{B_{B_s}} / f_{B_d} \sqrt{B_{B_d}}$ &$1.10 \pm 0.07$\cite{milc}\\
\hline
$A$  &  $0.81 \pm 0.04$  \\
\hline
$m_t(m_t)$  &  $168 \pm 6$ GeV  \\
\hline
\end{tabular}
\end{center}
\end{table}

\section{Pure superweak theories} \label{sec:psw}

A superweak theory \cite{sw} is one in which the CKM matrix is real,
so $\eta = 0$, and  ${\cal H}^{\Delta S = 2}_{eff}$ of eq. (1) originates
from physics outside the standard model. We define a {\it pure}
superweak theory to be one where all flavor changing phenomena (other
than $\epsilon_K$) are accurately described by the real CKM matrix.
%The dotted lines of Figure 1 show the $\chi^2$ contours for a fit with
%just the two observables $|V_{ub}/V_{cb}|$ and $\Delta M_{B_d}$.
Comparing Figures 1 and 2 at low $\eta$, one sees that the new limit
on $B_s$ mixing has excluded superweak theories with negative $\rho$.
This has important phenomenological consequences for
pure superweak theories.

We have computed $\chi^2(\rho)$ in pure
superweak theories, using as input the three observables
$|V_{ub}/V_{cb}|$, $\Delta M_{B_d}$ and $\Delta M_{B_s}$. We find that
all negative values of $\rho$ are excluded at greater than 99\%
confidence level. At positive $\rho$ only the two observables
$|V_{ub}/V_{cb}|$ and $\Delta M_{B_d}$, are relevant, and we find
the most probable value of $\rho$ to be +0.27.
However, even this value of $\rho$ corresponds to the pure
superweak theory being excluded at 92\% confidence level. Since the
uncertainties are dominated by the theory of $f_{B_d} \sqrt{B_{B_d}}$,
we take the view
that this does not  exclude purely superweak theories. In such
theories positive values of $\rho$ are 40 times more probable than negative 
values, and hence large values for
$f_B \sqrt{B_B} \approx 250$ MeV and small values for $|V_{ub}/V_{cb}|
\approx 0.06$ are predicted. 
A pure superweak description of $CP$ violation implies
\begin{equation}
+0.20 \;(0.13) < \rho < 0.34 \;(+0.41) \hspace{1in}
\mbox{at 68\% (95\%) confidence level}
\label{eq:rho}
\end{equation}

An important consequence of the new limit on $B_s$ mixing is the strong
preference for positive $\rho$ and the resulting small values for
$|V_{td}| \propto 1-\rho$. This is numerically significant: without
the $B_s$ mixing result the superweak theory can also have negative
values of $\rho$ which give $|V_{td}|$ about a factor of two larger
than the positive $\rho$ case.
With the $B_s$ result, a pure superweak theory must have $|V_{td}|$ at
the lower end of the standard model range. Thus in a pure superweak
theory, $\Delta M_{B_s} \propto \Delta M_{B_d} / |V_{td}|^2$ is
predicted to be
\begin{equation}
14 \;(10) \; \mbox{ps}^{-1}  < (\Delta M_{B_s})_{PSW} < 26 \; (32) \;
\mbox{ps}^{-1} \hspace{0.5in} \mbox{at 68\% (95\%) confidence level}
\label{eq:bssw}
\end{equation}
By comparison, in the standard
model $10.5 \; (9.5) \; \mbox{ps}^{-1} < \Delta M_{B_s} <
15 \; (19) \;\mbox{ps}^{-1}$ at 68\% (95\%) confidence level.

In the standard model, the branching ratio
$B(K^+ \rightarrow \pi^+ \nu \bar{\nu})$ is given by \cite{buras}
\begin{equation}
B(K^+ \rightarrow \pi^+ \nu \bar{\nu}) =
 c_1\left( (c_2 + c_3 A^2 (1 - \rho))^2
+ (c_3 A^2 \eta)^2 \right)
\label{eq:kpism}
\end{equation}
where $c_1 = 3.9 \times 10^{-11}, c_2 = 0.4 \pm 0.06$ and 
$c_3 = 1.52 \pm 0.07$.
In pure superweak theories, since $\rho$ is positive and $\eta=0$,
the branching ratio is
lowered to
\begin{equation}
B(K^+ \rightarrow \pi^+ \nu \bar{\nu}) = (5.0 \pm 1.0) \cdot 10^{-11}
\label{eq:kpisw}
\end{equation}
relative to the standard model prediction of $(6.6^{+1.4}_{-1.2})
\cdot 10^{-11}$.\footnote{This standard model result is smaller than that
quoted in the literature because the improved limit on $B_s$ mixing increases
$\rho$ even in the standard model.}
The recent observation of a candidate event for this decay \cite{ktopi}
is not
sufficient to exclude pure superweak theories, but further data from
this experiment could provide evidence against such theories.

\section{General superweak theories} \label{sec:gsw}

Pure superweak theories are artificial: they do not possess a symmetry which
allows ${\cal H}^{\Delta S = 2}_{eff}$ of eq. (1), while forbidding
similar $\Delta B = 2$ operators. If $\epsilon_K$ is generated by new
physics, why does this new physics not contribute to $B \bar{B}$
mixing? In general it would be expected to also contribute to  $\Delta
S = 1$ and  $\Delta B = 1$ processes. In the absence of a fundamental
theory of flavor, the relative sizes of the various flavor changing
operators can be estimated only by introducing arguments based on
approximate flavor symmetries.

We assume that the underlying theory of flavor possesses a flavor
symmetry group, $G_f$, and a mass scale $M_f$. The breaking of $G_f$,
whether explicit or spontaneous, is described in the low energy
effective theory by a set of dimensionless parameters, \{$\epsilon$\},
each with a well defined $G_f$ transformation. The low energy effective
theory of flavor is taken to be the most general operator expansion in
powers of $1/M_f$ allowed by $G_f$ and \{$\epsilon$\}. In the case
that the CKM matrix can be made real, we call these {\it general}
superweak theories. The phenomenology of such theories depends on
$G_f$, $M_f$ and \{$\epsilon$\} and will typically not coincide with the pure
superweak phenomenology. The $\Delta B = 2$ operators may lead to
exotic CP violation in neutral $B$ meson decays and may contribute to
$\Delta M_{B_d}$, allowing large values of $|V_{td}|$ invalidating
(\ref{eq:bssw}). Similarly the $\Delta S = 1$ operators  may invalidate
(\ref{eq:kpisw}), and may give an
observable contribution to $\epsilon'/\epsilon$.

\section{The effective Hamiltonian for the ``3 mechanism''} \label{sec:3ssw}

The dominant flavor changing neutral current (FCNC) interactions of the
down sector of the standard
model result from the ``3 mechanism'':
small flavor breaking parameters which mix the light quarks with the
heavy third generation quarks, together with a large, order unity,
breaking of the flavor symmetry that distinguishes the third generation
from the first two.
Hence, beneath the weak scale, the standard model yields an effective
Hamiltonian with dominant FCNC operators which contain a factor
$V_{ti}^* V_{tj}$ for each flavor changing current 
$\bar{d}_i d_j$, and a factor
$G_F^2 m_t^2/16 \pi^2 \approx (1/16 \pi^2) (1/v^2)$ from the loop integration.
The relevant diagrams are all 1 loop, giving the $(1/16 \pi^2)$ factor, and
involve the large GIM violation of the top quark mass; since there is no
small flavor violating parameter, the rest of the loop integral has an order
of magnitude given by dimensional analysis as $(1/v^2)$.

Now consider physics beyond the standard model where
the entire flavor structure of the theory beneath $M_f$ is controlled by $G_f$
and \{$\epsilon$\} --- both the Yukawa matrices of the standard model,
$\lambda(\epsilon)$, and the non-standard model operators in ${\cal H}_{eff}
(\epsilon)$. Since the dominant down sector, FCNC effects from $\lambda
(\epsilon)$ are known to arise from the ``3 mechanism'', we assume that $G_f$
and \{$\epsilon$\} are chosen so that the dominant such effects from
${\cal H}_{eff}(\epsilon)$ are also from the ``3 mechanism''.

The most general parameterization of the ``3 mechanism'' in the down sector
involves four complex parameters: $\epsilon_{L_i} = |\epsilon_{L_i}|
e^{i \phi_{L_i}}$ and $\epsilon_{R_i} = |\epsilon_{R_i}|
e^{i \phi_{R_i}}$, $i=1,2$, which describe the mixing of $d_{L_i}$ and
$d_{R_i}$ with $b_L$ and $b_R$. Assuming all phases to be of order unity, we
can describe the ``3 mechanism'' in terms of just four real small parameters
$|\epsilon_{L_i}|$ and $|\epsilon_{R_i}|$. We make the additional simplifying
assumption that $|\epsilon_{L_i}| = |\epsilon_{R_i}| =
\epsilon_i$, yielding the non-standard model interactions\footnote{It is
straightforward to extend this Hamiltonian to the most general case of the
``3 mechanism'' involving four complex parameters.}
\begin{eqnarray}
{\cal H}_{eff}^{(3)} & =  {1 \over M_f^2} & [
C_1 \; (\epsilon_1 \epsilon_2)^2 \; (\bar{s} d)^2 +
C_2 \; \epsilon_2^2 \; (\bar{s} b)^2 		  +
C_3 \; \epsilon_1^2 \; (\bar{b} d)^2              \nonumber \\
& &+ C_{1l} \; \epsilon_1 \epsilon_2 \; (\bar{s} d)(\bar{l} l) +
C_{2l} \; \epsilon_2 \; (\bar{s} b)(\bar{l} l) +
C_{3l} \; \epsilon_1 \; (\bar{b} d)(\bar{l} l) + \ldots ]
\label{eq:h3}
\end{eqnarray}
where $C_i$ are complex coefficients of order unity, and $l$ is a
lepton field.\footnote{We do not consider lepton flavor violation in
this letter.} A sum on possible gamma matrix structures is understood
for each operator. Since the flavor changing interactions from both the 
standard model and the new physics are governed by the same symmetry, we can 
choose $\epsilon_1 = |V_{td}|$ and $\epsilon_2 = |V_{ts}|$.
Such interactions can arise from many choices of $G_f$ and \{$\epsilon$\}; the
particular choice is unimportant, however, as the phenomenology rests
only on three assumptions
\begin{itemize}
\item There is an underlying theory of flavor based on symmetry $G_f$ and
breaking parameters \{$\epsilon$\}.
\item The dominant non-standard model FCNC operators of the down sector
arise from the ``3 mechanism''.
\item The symmetry breaking parameters of the down sector are
left-right symmetric, and have phases of order unity.
\end{itemize}

In the standard model, the dominant FCNC of the down sector arises
from the ``3 mechanism'', so that it is useful to describe the
effective theory beneath the weak scale by eq. (\ref{eq:h3}) with
\begin{equation}
\epsilon_1 = V_{td} \hspace{.5in} \epsilon_2 = V_{ts}^* \hspace{.5in}
{1 \over M_f^2} = {1 \over 16 \pi^2} {1 \over v^2}
\label{eq:h3sm}
\end{equation}
and $C_i$ real. This special case of the ``3 mechanism'' has a
restricted set of gamma structures due to the left-handed nature of
the weak interaction.

\section{Phenomenology of the ``3 mechanism'' in superweak theories}
\label{sec:p3m}
We have argued that pure superweak theories are artificial, and we now
study superweak theories where FCNC interactions are generated by the
``3 mechanism'' and yield ${\cal H}_{eff}^{(3)}$ of (\ref{eq:h3}). Why
should such theories have $V_{ij}$ real when $C_i$ are complex? One
possibility is that $G_f$ forces the Yukawa matrices $\lambda(\epsilon)$
to have a sufficiently simple form that they can be made real by field
redefinitions. Another possibility will be discussed later.

Since ${\cal H}_{eff}^{(3)}$ will be the origin of all $CP$ violation,
one may wonder if it could also account for all of
$\Delta M_{B_{d,s}}$. This is not possible --- charged current
measurements, together with the unitarity of $V$, imply $|V_{td}|$ and
$|V_{ts}|$ are sufficiently large that $W$ exchange contributes a
significant fraction of  $\Delta M_{B_{d,s}}$.

Given that the FCNC of both the standard model and exotic interactions have the
form of (\ref{eq:h3}), it would appear that the exotic interactions
must give a large fraction of $\Delta M_{B_{d,s}}$ since they are
responsible for all of $\epsilon_K$. This is not the case; in the
standard model the $\Delta S=2$ and $\Delta B = 2$ operators have
chirality $LL$, whereas for a generic ``3 mechanism'' they will have
all chiral structures. It is known that the $LR$, $\Delta S = 2$,
operator has a matrix element which is enhanced by about an order of
magnitude relative to that of the $LL$ operator \cite{latt},
and that there is no
similar enhancement in the $\Delta B = 2$ case. Furthermore, the
$LR$ operator is enhanced by QCD radiative corrections
in the infrared \cite{bagger}; with the
enhancement at 1 GeV about a factor of 3 larger than at 5 GeV. Hence
we conclude {\it In a generic superweak theory, we expect that}
${\cal H}_{eff}^{(3)}$ {\it leads to $\approx 3 \%$ contributions to
$\Delta M_{B_{d,s}}$.} There is considerable uncertainty in this
percentage because of the uncertainty in the overall enhancement of the
$\Delta S = 2$ and $\Delta B = 2$ contributions from the $LR$ operator,
and because of the unknown order unity $C_i$
coefficients. Given this result, we must evaluate how well these
generic superweak theories can account for the data, and to what
extent they lead to predictions.

Let $\Delta_{d,s}$ and $\delta_{d,s}$ be the standard model and new
physics contributions to
\begin{equation}
\Delta M_{B_{d,s}} = \Delta_{d,s} + \delta_{d,s}
\label{eq:db}
\end{equation}
First we consider a perturbation around the pure superweak case, where the
fractional contributions from new physics $F_{d,s} =
\delta_{d,s} / \Delta M_{B_{d,s}}$ are small. The central value of $\rho$,
from $\Delta M_{B_d}$ alone, changes by $\Delta \rho = 0.5 F_d$
for very small
$F_d$ ($\Delta \rho \simeq 0.3 F_d$ for $F_d \simeq 0.1$).
For positive $F_d$, this
improves the fit of general superweak theories to $\Delta M_{B_d}$ and
$|V_{ub}/V_{cb}|$. For example, $F_d = 0.1$ gives a central value of
$\rho= 0.28$ with
$\chi^2(\rho=0.28) \simeq 2.4$, which corresponds to 68\% C. L.
Since the allowed range of $\rho$ is little changed from eq. (5),
the prediction of small $|V_{td}|$ persists in these general superweak
theories, so that the prediction of eq. (7) for low values of
$B(K^+ \rightarrow \pi^+ \nu \bar{\nu})$ applies. Similarly, since $\rho$
is little altered, the prediction for $B_s$ mixing is $\Delta M_{B_{s}}
= (\Delta M_{B_{s}})_{PSW} (1 - F_d + F_s)$, where the pure superweak
prediction $(\Delta M_{B_{s}})_{PSW}$ is given in eq. (6). In this case
the general superweak theory also predicts large values of $\Delta
M_{B_{s}}$, although for negative $F_s$, it is not quite so large as
$(\Delta M_{B_{s}})_{PSW}$.

There is a second class of general superweak theories which is not a
perturbation about the parameters of the pure superweak theories. In
general superweak theories, the limit $\Delta M_{B_{s}} > 10.2$ ps$^{-1}$
can be expressed as $\rho > -0.06 + 0.5(F_d - F_s)$. For negative $F_d$ and
positive $F_s$, the negative $\rho$ region could become allowed. For
example, $F_s = -F_d = 0.1 \; (0.05)$ gives a theory
in which $\rho$ has a probability 25\% (9\%) of being negative.
This class of superweak theories {\it requires} values of $|F_{d,s}|$ which
are larger than our expectation, and appear somewhat improbable.
They have $|V_{td}|$ and $B(K^+ \rightarrow \pi^+ \nu \bar{\nu})$ at
the upper end of the standard model range. In these theories
$\Delta M_{B_{s}}$ is likely to be low, although it depends on $F_{d,s}$.

\section{Supersymmetry with a ``3 mechanism''} \label{sec:susy}

In general, the alternative theory of $CP$ violation of ${\cal H}_{eff}^{(3)}$
from the ``3 mechanism'' is not a strong competitor to the CKM theory
of $CP$ violation. The CKM theory, with two small measured
parameters, $|V_{us}|$ and $|V_{cb}|$, yields the correct order of
magnitude for $\epsilon_K$, while superweak theories with the ``3
mechanism'' apparently require a new scale $M_f \approx 30 v
\approx 10$ TeV. However, there is the interesting possibility that
the new physics generates FCNC operators only at 1 loop, as in the
standard model. This would give $M_f \approx 4 \pi m_f$, with the mass
of the new quanta close to the weak scale at $m_f \approx 1$ TeV. We
therefore take the view that the ``3 mechanism'' generating FCNC
operators at 1 loop at the weak scale is a credible alternative to the
CKM theory of CP violation. While not as minimal as the CKM theory, it
correctly accounts for the order of magnitude of $\epsilon_K$.

Let $l$ represent $d,s$ or $b$, left or right handed. New interactions
of the form $\bar{l}lH$, where $H$ is some new heavy field, will generate
FCNC at tree level, whereas $lHH$ generates them at 1 loop. Thus the
exotic new heavy particles at the weak scale should possess a parity
so that they appear only in pairs.

Weak scale supersymmetry allows a symmetry description of the weak
scale, and leads to a successful prediction for the weak mixing
angle. Furthermore, it incorporates the economical Higgs description of
flavor of the standard model. $R$ parity ensures that superpartners
appear pairwise in interactions, so that the dominant supersymmetric
contributions to FCNC processes occur only at one loop. Supersymmetric
theories have several new generation mixing matrices --- in particular
$W_{L,R}$ at the gluino interaction $(\tilde{d}^\dagger_{L,R} W_{L,R}
d_{L,R} )\tilde{g}$. A flavor symmetry, $G_f$, can ensure that the largest
contribution from superpartner exchange to FCNC occurs via the ``3
mechanism''\cite{bdh,bhr}.
If the small symmetry breaking parameters are left-right
symmetric and real, this gives ${\cal H}^{(3)}_{eff}$ of (\ref{eq:h3}) with
\begin{equation}
|W_{{L,R}_{31}}| \approx \epsilon_1 = |V_{td}| \hspace{.5in}
|W_{{L,R}_{32}}| \approx \epsilon_2 = |V_{ts}| \hspace{.5in}
{1 \over M_f^2} = {1 \over 16 \pi^2} {1 \over \tilde{m}^2}
\label{eq:h3ss}
\end{equation}
where $\tilde{m}$ is the average mass of the colored superpartners in
the loop. As the superpartners are at the weak scale, $\tilde{m}
\approx v$, and comparing with  (\ref{eq:h3sm}) one finds that,
with weak scale supersymmetry, it may well be that
$\epsilon_K$ receives comparable standard model and supersymmetric
contributions.\footnote{Given the order of magnitude enhancement of the
matrix element of the $LR$ operator relative to the $LL$, and given the
further order of magnitude enhancement of $C_{LR}$ relative to  $C_{LL}$
from QCD scaling, one generically expects the supersymmetric contribution to
be larger. However, these factors may be outweighed by colored superpartner
masses somewhat larger than $v$,
some degree of degeneracy between the third generation scalars and those of
the lighter generations, and by $W_{ij}$ somewhat less than $V_{ij}$.
We note that the QCD enhancement of $C_{LR}$ for the $\Delta S = 2$ 
operator\cite{bagger} was not included in \cite{bdh,bhr,bs}}

Here we stress that weak scale supersymmetry can provide an important
example of the general superweak theories discussed in this
letter. The absence of CKM $CP$ violation would be
guaranteed if $CP$ violation were {\it soft} --- restricted to operators of
dimension two and three. The Yukawa matrices would then be real, so that there
would be no $CP$ violation from diagrams with internal quarks, but the
scalar mass matrices would contain phases, so that $CP$ violation would
arise from diagrams with internal squarks.\footnote{This is an alternative
view
to the one presented in \cite{bhr}, where the specific flavor symmetry forces
forms for $V$ and $W$ matrices such that even the supersymmetric contribution
to $\epsilon_K$ involves a phase originating from the Yukawa couplings.}
Soft $CP$ violation in
supersymmetric theories, with FCNC operators arising from the ``3
mechanism'', represents a well-motivated and credible alternative
to CKM $CP$ violation, and will be explored in detail elsewhere.

%\begin{equation}
%
%\label{eq:}
%\end{equation}

\section{Summary} \label{sec:sum}

Fits of the CKM matrix to
$|V_{ub}/V_{cb}|$, $\Delta M_{B_d}$ and $\Delta M_{B_s}$
show that at 68\% C.L. the standard model correctly predicts $\epsilon_K$
to better than a factor of two, while at 90\% C.L. not even the order
of magnitude can be predicted. On one hand the standard model is
highly successful; on the other, there is still room for an
alternative theory of CP violation.

The recent improvement on the limit on $\Delta M_{B_{s}}$\cite{jeru}
implies that pure superweak theories with negative $\rho$ are excluded, while
at positive $\rho$ they are somewhat disfavored. Pure superweak theories
allow $0.13 < \rho < 0.41$ at 95\% C.L., and predict high values
for $\Delta M_{B_{s}}$ and $f_B \sqrt{B_B}$ and low values for
$|V_{ub}/V_{cb}|$, $B(K^+ \rightarrow \pi^+ \nu \bar{\nu})$ and
$\epsilon'/\epsilon$.

We have argued that {\it pure} superweak theories are artificial, and
have introduced {\it general} superweak theories, in which all FCNC are
governed by an approximate flavor symmetry and the ``3 mechanism.'' In
this case the new physics induces other
flavor changing operators in addition to the
$\Delta S = 2$ operator responsible for $\epsilon_K$; in particular,
${\cal O} (3)\%$ contributions to $B_{d,s}$ mixing are expected.
There are two important
classes of general superweak theories, one with positive $\rho$ and the
other with negative $\rho$.
The first can be viewed as a perturbation about the superweak case, with
an improved fit to data, while retaining the characteristic predictions
mentioned above. The negative $\rho$ possibility appears less likely, and
arises only if the new
physics contributes more than 10\% of $\Delta M_{B_{d,s}}$. In
this case future data should show a high value for
$B(K^+ \rightarrow \pi^+ \nu \bar{\nu})$ and low values for $\Delta M_{B_{s}}$,
$f_B \sqrt{B_B}$, $|V_{ub}/V_{cb}|$, and $\epsilon'/\epsilon$.
All these superweak theories predict low values for the $CP$ asymmetries in
$B$ meson decays.

Weak scale supersymmetric theories with softly broken  $CP$ can
provide an important example of general superweak theories. As in the CKM
theory, assuming phases of order unity yields a 
correct prediction for the order of magnitude of $\epsilon_K$. In
addition they have $\bar{\theta}=0$ at tree level, and it is interesting
to seek a flavor symmetry which would sufficiently protect $\bar{\theta}$
from radiative corrections to solve the strong $CP$ problem.

\begin{center}
{\bf Acknowledgments}
\end{center}
%{\em This work was supported by the Director, Office of Energy Research,
%Office of High Energy and Nuclear Physics, Division of High Energy
%Physics of the U.S. Department of Energy under Contract DE-AC03-76SF00098.}
This work was supported in part by the Director, Office of
Energy Research, Office of High Energy and Nuclear Physics, Division of
High Energy Physics of the U.S. Department of Energy under Contract
DE-AC03-76SF00098 and in part by the National Science Foundation under
grant PHY-95-14797.

%\appendix

% '99' is at least as wide as the widest bibliography label,
% could use '9' if there are less than 10 references.

%\newpage

%\begin{center}

%{\bf Figure Captions}
%\end{center}

%{\bf Fig. 1}   Diagrammatic representation of the operators generated by
%heavy particle exchange.  The operators shown contribute to the up
%and down quark Yukawa matrices when the flavons acquire vacuum expectation
%values.

\end{document}